\documentclass[structuredabstract]{aa}
\usepackage{graphicx}
\usepackage{natbib}
\usepackage{txfonts}

\begin{document}
 
\title{Cooling of young neutron stars in GRB associated to supernovae}

\author{Rodrigo Negreiros\inst{1}, Remo Ruffini\inst{2,3,4}, Carlo
Luciano Bianco\inst{2,3}, \and Jorge A. Rueda\inst{2,3}}

\institute{Frankfurt Inst. for Adv. Studies, Johann Wolfgang Goethe
University, Ruth-Moufang-Str. 1, 60438 Frankfurt am Main, Germany.
\and
Dipartimento di Fisica and ICRA, Universit\`a di Roma ``La Sapienza'', Piazzale Aldo Moro 5, I-00185 Roma, Italy.
\and
ICRANet, Piazzale della Repubblica 10, I-65122 Pescara, Italy.
\and
ICRANet, Universit\'e de Nice Sophia Antipolis, Grand Ch\^ateau, BP 2135, 28, avenue de Valrose, 06103 NICE CEDEX 2, France.
}

\authorrunning{R. Negreiros et al.}

\titlerunning{Cooling of young neutron stars in GRB associated to SN}

\date{}

\abstract
{
Recent observations of the late ($t=10^8$--$10^9$ s) emission of the supernovae (SNe) associated to GRBs (GRB-SN) show a distinctive emission in the X-ray regime consistent with temperatures $\sim 10^7$--$10^8$ K. Similar features have been also observed in two Type Ic SNe SN 2002ap and SN 1994I that are not associated to GRBs.
}
{
We advance the possibility that the late X-ray emission observed in GRB-SN and in isolated SN is associated to a hot neutron star just formed in the SN event, here defined as a neo-neutron star.
}
{
We discuss the thermal evolution of neo-neutron stars in the age regime that spans from $\sim 1$ minute (just after the proto-neutron star phase) all the way up to ages $<$ 10--100 yr. We examine critically the key factor governing the neo-neutron star cooling with special emphasis on the neutrino emission. We introduce a phenomenological heating source, as well as new boundary conditions, in order to mimic the high temperature of the atmosphere for young neutron stars. In this way we match the neo-neutron star luminosity to the observed late X-ray emission of the GRB-SN events: URCA-1 in GRB980425-SN1998bw, URCA-2 in GRB030329-SN2003dh, and URCA-3 in GRB031203-SN2003lw.
}
{
We identify the major role played by the neutrino emissivity in the thermal evolution of neo-neutron stars. By calibrating our additional heating source at early times to $\sim 10^{12}$--$10^{15}$ erg/g/s, we find a striking agreement of the luminosity obtained from the cooling of a neo-neutron stars with the prolonged ($t=10^{8}$--$10^{9}$ s) X-ray emission observed in GRB associated with SN. It is therefore appropriate a revision of the boundary conditions usually used in the thermal cooling theory of neutron stars, to match the proper conditions of the atmosphere at young ages. The traditional thermal processes taking place in the crust might be enhanced by the extreme high-temperature conditions of a neo-neutron star. Additional heating processes that are still not studied within this context, such as $e^+e^-$ pair creation by overcritical fields, nuclear fusion, and fission energy release, might also take place under such conditions and deserve further analysis.
}
{
Observation of GRB-SN has shown the possibility of witnessing the thermal evolution of neo-neutron stars. A new campaign of dedicated observations is recommended both of GRB-SN and of isolated Type Ic SN. 
}

\keywords{Stars: neutron --- Gamma-ray burst: general --- supernovae: general}

\maketitle

\section{Introduction}

The investigation of the thermal evolution of neutron stars is a powerful tool for probing the inner composition of these objects. The cooling of neutron stars has been investigated by many authors, where many different microscopic models were assumed \citep[see][]{Schaab1996,Page2004,Page2006,Page2009,Blaschke2000, Grigorian2005,Blaschke2006,Negreiros2010}. Most of the research on the thermal evolution of compact stars focus on objects older than 10 to 100 years, which is comprehensible if one considers that the thermal data currently available to us is for pulsars with estimated ages the same as or greater than 330 years \citep{Page2004,Page2009}. We discuss the thermal evolution of young neutron stars, in the little explored time window that spans ages greater than 1 minute (just after the proto-neutron star regime \citep{Prakash2001}) to ages $\leq 10$--$100$ years, when the neutron star becomes isothermal \citep[see][for details]{Gnedin2001}. 

We discuss the possibility that the late X-ray emission (URCA hereafter \footnote{The names URCA-1 and URCA-2 mentioned here were given to these sources when presented for the first time at the MG10 meeting held in Rio de Janeiro in the town of Urca. The location of the MG10 meeting was very close to the ``Cassino da Urca'' where George Gamow and Mario Schoenberg conceived the process of neutrino emission for the cooling process of neutron stars, which also took the name from the town of Urca, the Urca process \citep[see e.g detailed history in][]{2005tmgm.meet..369R,1970mwla.book.....G}}) following a few GRBs associated with supernovae (SNe); e.g. URCA-1 in GRB980425-SN1998bw \citep{2004AdSpR..34.2715R,2005tmgm.meet.2451F,GraziaMG11}, URCA-2 in GRB030329-SN2003dh \citep{2004AIPC..727..312B,2005tmgm.meet.2459B}, and URCA-3 in GRB031203-SN2003lw \citep{2005ApJ...634L..29B,2007ESASP.622..561R,2008ralc.conf..399R} (see Fig.~\ref{fig:xray_lc} for details), might actually be created from young ($t \sim$ 1 minute--$(10$--$100)$ years), hot ($T \sim 10^7$--$10^8$ K) neutron stars that are remnants of the SN \citep{2007ESASP.622..561R} and which we here call neo-neutron stars. Relevant also are the observations of the isolated Type Ic supernova SN 1994I \citep{2002ApJ...573L..27I} and SN 2002ap \citep{2004A&A...413..107S}, which present late emissions similar to the ones observed in URCA-1, URCA-2, and URCA-3.

Here we propose a revision of the boundary conditions usually employed in the thermal cooling theory of neutron stars, in order to match the proper conditions of the atmosphere at young ages. We also discuss the importance of the thermal processes taking place in the crust, which also have strong effects on the initial stages of thermal evolution. We stress that we are not calling the validity of the current treatment of the atmosphere of compact stars into question but, instead, we point out the need of extending them to appropriately describe the conditions of neo-neutron stars.

\section{Cooling of young, hot neutron stars}

\begin{figure}
\begin{center}
 \includegraphics[width=\hsize,clip]{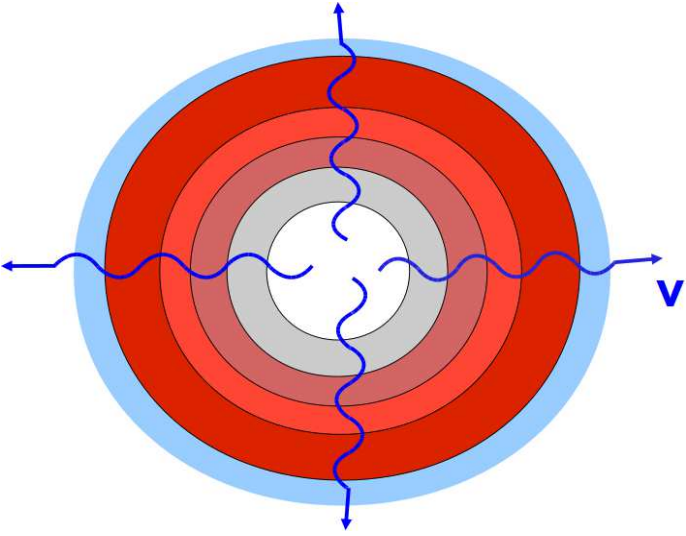}
\end{center}
\caption{Schematic representation of the cooling of a young neutron star. Due to stronger neutrino emissivities, the core of the star cools down more quickly than the crust, causing the star to cool inside out. Darker and lighter areas
represent higher and lower temperatures, respectively.}
\label{fig:cool_scheme} 
\end{figure}

There are three important ingredients that govern the thermal evolution of a compact star: 1) the microscopic input, which accounts for the neutrino emissivities, specific heat, and thermal conductivity; 2) the macroscopic structure of the star, namely its mass, radius, pressure profile, crust size, etc.; and 3) the boundary condition on the surface of the star, which provides a relationship between the mantle temperature and that of the atmosphere, the latter being what we ultimately observe. These ingredients have been studied extensively, and a comprehensive review can be found in \citet{Page2006}. As discussed in \citet{Gnedin2001}, during the initial stages of thermal evolution (ages $\leq 10 -100$ years), the core and the crust of the neutron star are thermally decoupled. This is because the high-density core is emitting neutrinos at a much higher rate than the crust, which causes it to cool down more quickly. This effectively means that, initially, the neutron star is cooling ``inside out'', with the core colder than the outer layers. This scenario is schematically depicted in Figure \ref{fig:cool_scheme}.

The dominant neutrino emission processes in the crust are given by the Bremsstrahlung, plasmon decay, and electron-positron annihilation processes. Following the footsteps of \citet{Gnedin2001}, we calculate the thermal evolution of neutron stars, by artificially adding a phenomenological source of heat (see details in Sec.~\ref{sec:4}). This allow us to estimate how much heat is needed, so that the thermal evolution of a neo-neutron star matches the X-ray light curve of late emission of GRB-SN.

After this initial core-crust decoupled state, the ``cooling wave'' originated in the core reaches the crust, and the object becomes isothermal. The time scale of this process is between 10 and 100 years, depending on the properties of the crust \citep{Gnedin2001}. This means that the crust shields the core during the initial stages of thermal evolution, and all the information we might obtain at this stage refers only to the crust and to the atmosphere of the star. This raises another issue about the atmosphere of the star. The thermal connection between the mantle and the atmosphere is what defines the photon luminosity, which is what we observe. Therefore, the appropriate description of the atmosphere is key to correctly understanding the thermal evolution of neutron stars. In the usual approach, the thermal relaxation time of the atmosphere is assumed to be much shorter than that of the neutron star; furthermore, neutrino emissions from the atmosphere are also considered to be negligible \citep[see][]{Gudmundsson1983}. Under these assumptions, and assuming a plane-parallel approximation (which is reasonable since the atmosphere is $\sim 100$ m thick), one can get a relationship between the temperature of the mantle $T_b$ and the temperature of atmosphere $T_e$ or, equivalently, the luminosity $L_e$. \citet{Gudmundsson1983} originally found a $T_b$-$T_e$ relationship that depends on the surface gravity of the neutron star. This relationship was developed further by \citet{Potekhin1997} to account for the possibility of mass accreted in the initial stages and of magnetic fields effects. As pointed out by \citet{Gudmundsson1983}, such assumptions for the atmosphere of the star are only valid for objects older than a few tens of years, when the temperature has dropped below $10^9$K for densities below $10^{10}$ g/cm$^3$. In fact, we see that the current boundary conditions yields temperatures $\sim 10^7$ K ($L\sim 10^{37}$ erg/s, equivalently) for young neutron stars (age $<1$--$10$ years). This should raise some suspicion since proto-neutron stars studies \citep[see][and references therein]{Prakash2001} indicate that neutron stars just after this regime have temperatures $\sim ~ 10^{10}$--$10^{11}$ K.

\begin{figure}
\begin{center}
\includegraphics[width=\columnwidth,clip]{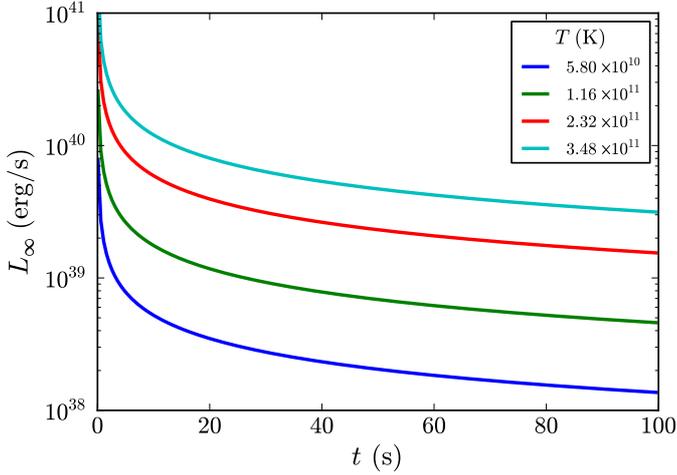}
\end{center}
 \caption{Luminosity of a hot nascent neutron star as observed at infinity given by Eq.~(\ref{eq:Lshapiro}) during the initial 100 s \citep{Duncan1986}, with the initial temperatures indicated. The neutron star is assumed to have a mass of $1.4 M_\odot$ and a radius of 13 km. 
\label{fig:L_shapiro}} 
\end{figure}

The properties of the atmosphere of a sufficiently hot, nascent neutron star should differ significantly from those considered in \citet{Gudmundsson1983} and \citet{Potekhin1997}, especially since at hot temperatures ($T \gtrsim 10^9$ K) the atmosphere might not be transparent to neutrinos, and thus the neutrino transport equations have to be considered. The coupled equations of neutrino and photon transport, in the atmosphere of a neutron star, were solved
by \citet{Salpeter1981}, and \citet{Duncan1986}. In these works the authors performed detailed calculations of the atmosphere properties of hot neutron stars. They found the following photon luminosity, as observed at infinity, 
\begin{equation}\label{eq:Lshapiro}
L_\infty =
50\times t^{-7/12}\times\left(T_{10}\right)^{7/4}
\times\left(R_{10}\right)^{17/9}\times
\left(\frac{M}{M_\odot}\right)^{-1} \times L_E, 
\end{equation}
where $t$ is time in seconds, $T_{10}$ is the initial temperature in units of 10 MeV, $R_{10}$ the neutron star radius in units of 10 km, $M$ the neutron star mass, and $L_E \sim 2.0 \times 10^{38}$ erg/s is the Eddington
luminosity. \citet{Duncan1986} found that the above expression should be valid for at least the initial 100 s. In Fig.~\ref{fig:L_shapiro} we can see how the luminosity of the star changes for the first 100 s, for stars with different initial temperatures.

According to these results, during the initial 100 s, the photon luminosity emerging from the atmosphere will be higher than the Eddington luminosity. This implies that there will be mass loss, owing to neutrino-driven winds from the young atmosphere. As shown by \citet{Duncan1986}, the total mass loss only becomes appreciable for neutron stars with large radii and high initial temperatures. For a typical neutron star with the canonical mass of $1.4 M_\odot$, a radius of 13 km and initial temperature of $\sim 10^{11}$ K, the total mass loss was estimated to be $\sim 6.2\times10^{-6} M_\odot$.

In addition to the high luminosities associated to the atmosphere of young neutron stars, one also needs to consider fallback onto the surface of the neutron star. \citet{Potekhin1997} discuss how fallback, at earlier
stages of evolution, would modify the properties of the atmosphere, hence of the boundary conditions. Once more in this investigation, however, such a fallback is assumed to have happened at early times, and the modified boundary conditions are only valid if the fallback has already ceased. \citet{Chevalier1989} studied the fallback onto young neutron stars and found that, while there is an envelope, a luminosity near the Eddington limit should be present. Furthermore, the authors have found that in this case the energy from the envelope can be radiated away in about one year. This timescale, however, might be lengthened if rotation effects are accounted for during the fallback. In addition to that, \citet{Turolla1994} discuss the possibility of ``hot solutions'' for the atmosphere of neutron stars undergoing spherical accretion. It is shown that for $L \geq 10^{-2}L_E$ the temperature at the atmosphere of a neutron star might be $\sim 10^9$--$10^{11}$ K.

\section{Late X-ray emission in GRBs associated to supernovae: URCAs}

\begin{figure}[h!]
\includegraphics[width=\hsize,clip]{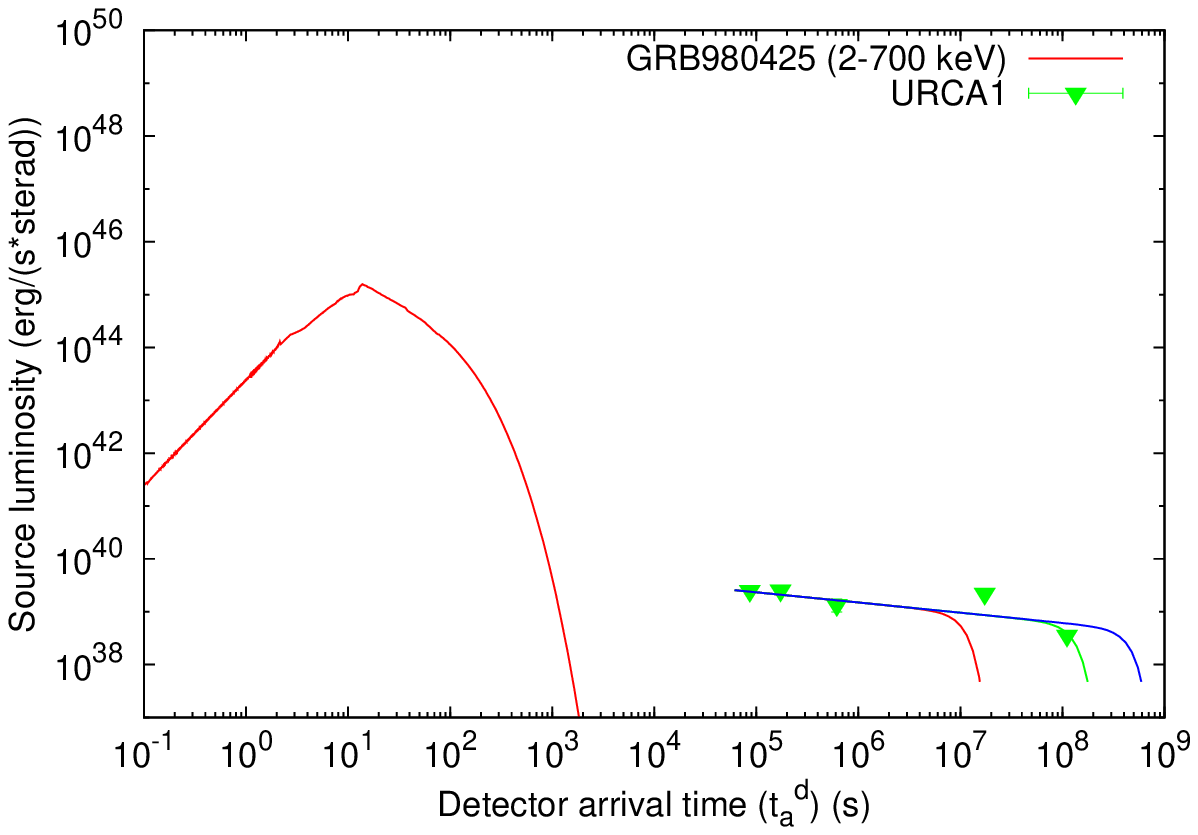}
\includegraphics[width=\hsize,clip]{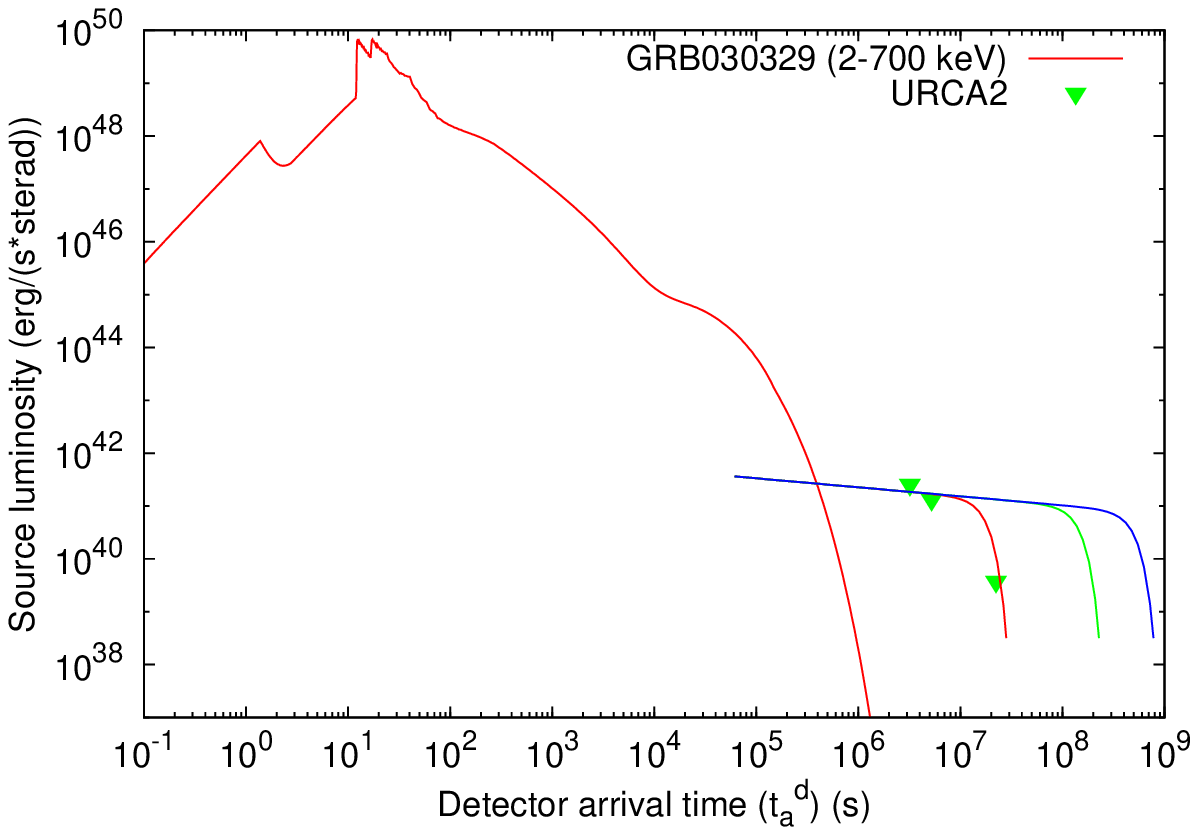}
\includegraphics[width=\hsize,clip]{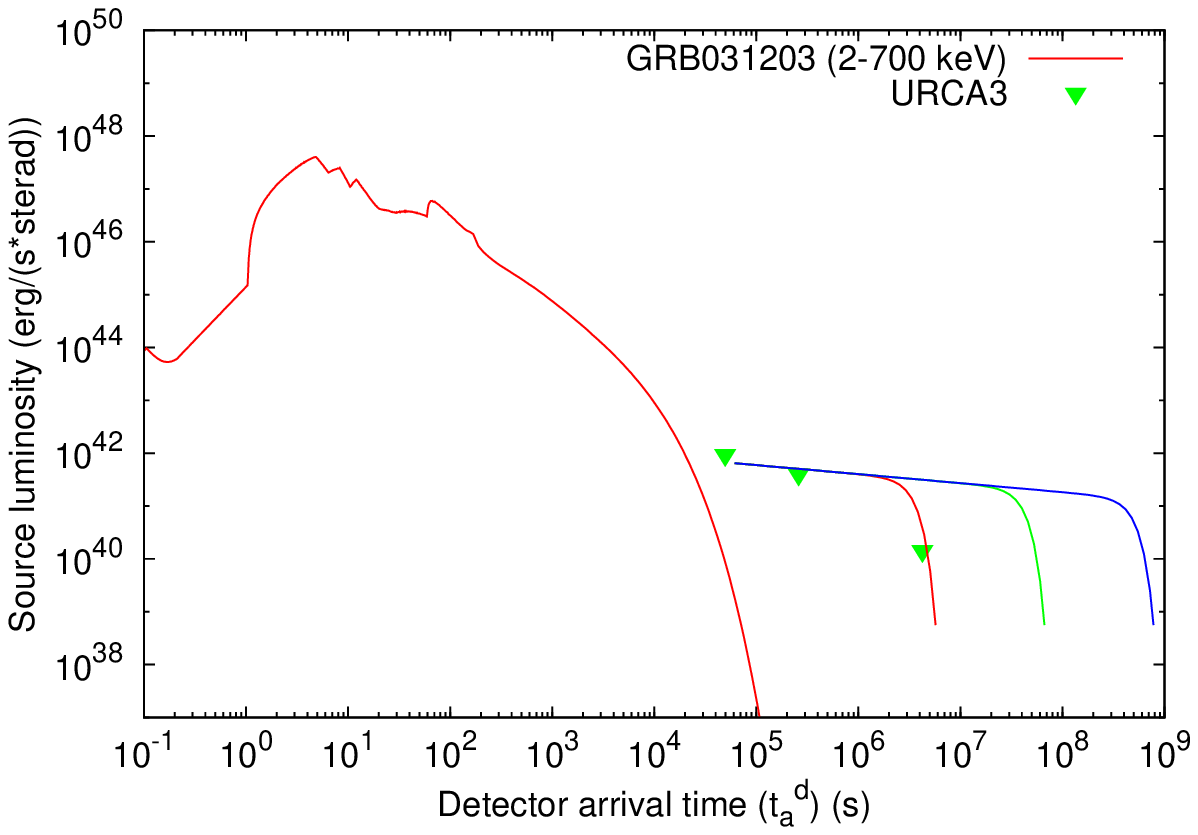}
\caption{Synthetic light curves of GRB980425 (A) \citep{2004AdSpR..34.2715R,2005tmgm.meet.2451F,GraziaMG11}
, GRB030329 (B) \citep{2004AIPC..727..312B,2005tmgm.meet.2459B}
and GRB031203 (C) \citep{2005ApJ...634L..29B,2007ESASP.622..561R,2008ralc.conf..399R}
. The solid curves represent the hard X-ray emission (10-200 keV range) and the triangles are 2-10 keV flux points. The optical luminosities of the SNe accompanying these GRBs are also reported with crosses \citep[see][for details]{2007ESASP.622..561R}. The curves fitting the late X-ray luminosity (URCAs) are qualitative cooling curves based on \cite{1978pans.proc..448C}; see also \cite{2004AdSpR..34.2715R,2007ESASP.622..561R,2008ralc.conf..399R}, \cite{2004AIPC..727..312B,2005ApJ...634L..29B,2005tmgm.meet.2459B,GraziaMG11}, and \cite{2005tmgm.meet.2451F} for details.}\label{fig:xray_lc} 
\end{figure}

It seems clear to us that, after the analysis of the scenario described above, we must extend the current model for the boundary conditions used in cooling calculations, to include the effects of a high-temperature atmosphere, possibly with super-Eddington luminosity. Up until this point,	 however, little attention has been given to the thermal evolution of young neutron stars, mainly due to the absence of observational data of neutron stars with ages $<330$
years. It has been recently proposed \citep[see][for details]{2007ESASP.622..561R} that the long-lasting X-ray emission called URCA there (see Fig.~\ref{fig:xray_lc}) of a few GRBs associated to SNe; URCA-1 in GRB980425-SN1998bw \citep{2004AdSpR..34.2715R,2005tmgm.meet.2451F,GraziaMG11}, URCA-2 in GRB030329-SN2003dh \citep{2004AIPC..727..312B,2005tmgm.meet.2459B}, and URCA-3 in GRB031203-SN2003lw \citep{2005ApJ...634L..29B,2007ESASP.622..561R,2008ralc.conf..399R}, might actually originate in the compact star remnant of the SN: a neo-neutron star. In this scenario the GRB is described as the core collapse of a massive star, whose remnant is a black hole. This massive star is supposed to be in a binary system, whose companion is on the verge of becoming an SN. The GRB triggers the SN explosion in the companion star, which in turns leaves behind a neutron star \citep{2001ApJ...555L.117R}. An alternative scenario has been recently suggested in which the so-called GRB is actually not a GRB but the observed X-ray emission stems from a collapsing core: a proto-neutron star leading directly to a SN explosion \citep{ruffinichina2011}. Such a process, if confirmed, will naturally explain the observations of the X-ray outburst 080109/SN 2008D \citep{2008Natur.453..469S}. This concept is very similar to the one of a proto-black hole as introduced in \cite{TEXAS,COSPAR,IZZOAA,2012A&A...538A..58P}, where the emission from the collapsing core is clearly distinguished from the GRB. In that case, the collapsing core leads to the formation of the black hole, while in the present case it forms a neutron star.

Both scenarios form a neo-neutron star, and they are supported by the observation of supernova 1979C \citep{Patnaude2011}, where a similar X-ray light curve also followed the SN. In Fig.~\ref{fig:xray_lc} we show the X-ray light curve associated with the URCAs.

\begin{figure}
\includegraphics[width=\hsize,clip]{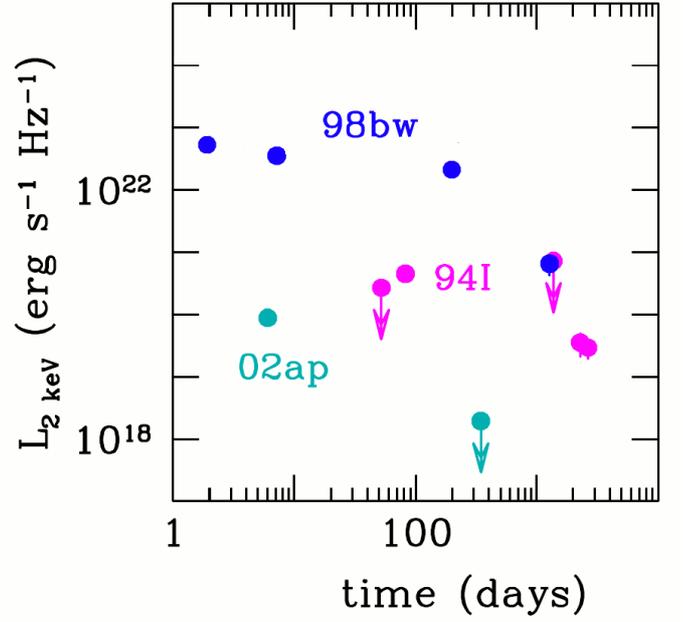}
\caption{X-ray light curves of the counterparts of GRB980425-SN1998bw and of two Type Ic SNe not accompanied by GRBs: SN1994I (``normal'') and SN2002ap (broad-lined). The data are from \citet{2000ApJ...536..778P}, \cite{2002ApJ...573L..27I}, \cite{2004ApJ...608..872K}, and \cite{2004A&A...413..107S}.}
\label{fig:fig_SN}
\end{figure}

From Fig.~\ref{fig:xray_lc} we can see that the X-ray luminosities of these sources are of the same magnitude as what is expected for neo-neutron stars, as discussed above. In Table \ref{tab:tabella} we summarize the representative parameters of the four GRB-SN systems, including the very high kinetic energy observed in all SNe \citep{mazzaliVen}. We have also included the association GRB060218-SN2006aj \citep[see][for details]{2007A&A...471L..29D,dainotti2010JKPS}. It must be noted that similar prolonged X-ray emission has also been observed in connection with other Type Ic SN not associated with GRBs, such as SN1994I \citep{2002ApJ...573L..27I} and SN2002ap \citep{2004A&A...413..107S} (see Fig.~\ref{fig:fig_SN} for details).

\begin{table*}
\centering
\caption{GRBs associated to SNe and URCAs}
{\footnotesize
\begin{tabular}{ccccccccc}
\hline
GRB & $\begin{array}{c}E_{e^\pm}^{tot}\\ \mathrm{(erg)}\end{array}$ & $\begin{array}{c}E_{SN}^{bolom}\\ \mathrm{(erg)^a}\end{array}$ & $\begin{array}{c}E_{SN}^{kin}\\ \mathrm{(erg)^b}\end{array}$ & $\begin{array}{c}E_{URCA}\\ \mathrm{(erg)^c}\end{array}$ & $\displaystyle\frac{E_{e^\pm}^{tot}}{E_{URCA}}$ & $\displaystyle\frac{E_{SN}^{kin}}{E_{URCA}}$ & $\begin{array}{c}R_{NS}\\ \mathrm{(km)^d}\end{array}$ & $z^e$ \\
\hline
980425 & $1.2\times 10^{48}$ & $2.3\times 10^{49}$ & $1.0\times 10^{52}$ & $3\times 10^{48}$ & $0.4$ & $1.7\times10^{4}$ & $ 8$ & $0.0085$\\
030329 & $2.1\times 10^{52}$ & $1.8\times 10^{49}$ & $8.0\times10^{51}$ & $3\times10^{49}$ & $6\times 10^{2}$ & $1.2\times10^{3}$ & $14$ & $0.1685$\\
031203 & $1.8\times 10^{50}$ & $3.1\times 10^{49}$ & $1.5\times10^{52}$ & $2\times10^{49}$ & $8.2$ & $3.0\times10^{3}$ & $20$ & $0.105$\\
060218 & $1.8\times 10^{50}$ & $9.2\times 10^{48}$ & $2.0\times10^{51}$ & $?$ & $?$ & $?$ & $?$ & $0.033$\\
\hline 
\end{tabular}}
\tablefoot{a) see \citet{2007ApJ...654..385K}; b) Mazzali, P., private communication at MG11 meeting in Berlin, July 2006, \cite{1998Natur.395..672I}; c) evaluated fitting the URCAs with a power law followed by an exponentially decaying part; d) evaluated assuming a mass of the neutron star $M=1.5 M_\odot$ and $T \sim 5$--$7$ keV in the source rest frame; e) see \citet{1998Natur.395..670G,2003GCN..2020....1G,2004ApJ...611..200P,2006ApJ...643L..99M}. Here $E^{tot}_{e^\pm}$ is the total energy of GRB, $E^{bolom}_{SN}$ and $E_{SN}^{kin}$ are the bolometric and the kinetic energy of the SN, $E_{URCA}$ is the energy of the late X-ray emission URCA (see Fig.~\ref{fig:xray_lc}), $R_{NS}$ is the radius of the neutron star, and $z$ is the redshift of the event.}
\label{tab:tabella}
\end{table*}

\section{Neo-neutron star luminosity and the URCAs}\label{sec:4}

Another important ingredient for the cooling of young neutron stars are the crust properties. As illustrated in Fig.~\ref{fig:cool_scheme}, due to the stronger neutrino emission from the core, the core and crust are thermally decoupled during the initial stages. For that reason, the initial stages of the thermal evolution reflect the properties of the crust, while the core remains invisible. Thus the proper description of the crust structure and composition is also fundamental for understanding the initial thermal stages in the evolution of a neutron star. We now briefly discuss the current understanding of the crustal processes and how it might be related to the data available from the URCAs.

There are several active emission mechanisms in the neutron star crust, e.g.~$e$-ion Bremsstrahlung, plasmon decay,  $e^+$-$e^-$ annihilation, $e$-$e$ and $n$-$n$ Bremsstrahlung, synchrotron emission, as well as Cooper pair processes for temperatures below the critical temperature for superfluidity $T_{\rm crit}$. However, as shown by \cite{Yakovlev2001a}, the first three processes are the dominant ones for temperatures above $10^8$ K, which is the regime we are interested in. For instance, synchrotron emission channels might become slightly relevant, but only for $T < 10^8$ K and for very high magnetic fields $>10^{14}$ G. The Cooper pair mechanism, possibly important for objects a few hundred years old like Cas A \citep[see e.g.][for details]{2011PhRvL.106h1101P,2011MNRAS.412L.108S}, is irrelevant in the present case since we are dealing with neutron star ages below ten years and thus temperatures well above $T_{\rm crit}$. 

At temperatures $T \sim 3\times 10^9$ K, we can write for the most important emission processes in the crust,
\begin{eqnarray}
 \epsilon_B \sim 10^{21} \textrm{erg s}^{-1}\textrm{cm}^{-3},\\
 \epsilon_P \sim 10^{22} \textrm{erg s}^{-1}\textrm{cm}^{-3},\\
 \epsilon_{ep} \sim 10^{19} \textrm{erg s}^{-1}\textrm{cm}^{-3},
\end{eqnarray}
where $\epsilon_i$ denotes the emissivity, and the indexes $B$, $P$, $ep$ denote the Bremsstrahlung, plasmon decay, and pair annihilation processes, respectively.

To estimate the amount of heat needed to match the theoretical thermal evolution of a neo-neutron star to the light curve of the URCAs, we added a phenomenological source of heat parametrized by
\begin{equation}\label{eq:H}
H = H_0\,e^{-t/\tau_S}\, ,
\end{equation}
with $H_0$ the magnitude of the heat source, and $\tau_S$ the time scale in which it is active. For our calculations we set $\tau_S = 1$ year.

In addition, we introduced a phenomenological boundary condition for the early stages of evolution of the surface temperature $T_s$ that follows the form $T_s = T_x\,gs_{14}^{1/4} T_8^{0.55}$ K, where $T_x =0.87\times 10^6  + (T_0-0.87 x 10^6)\,e^{-t/\tau_S}$ K, $T_8$ is the mantle temperature $T_b$ in units of $10^8$ K, $T_0$ is the initial temperature of the atmosphere, and $gs_{14}$ is the surface acceleration of gravity in units of $10^{14}$ cm/s$^2$. With this new boundary condition, we can mimic the high temperature of the atmosphere for young neutron stars by setting the temperature at early times to a higher value and, for times greater than $\tau_S$, it asymptotically goes to its traditional value $\sim 0.87\times 10^6$ K.

\begin{figure}
\includegraphics[width=\hsize,clip]{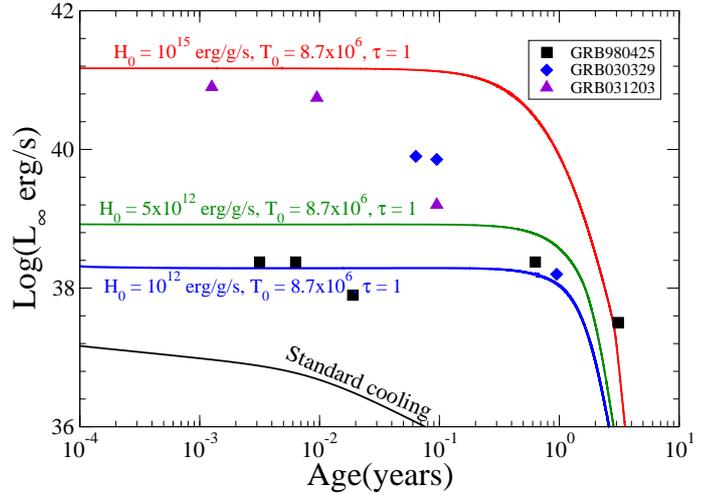}
\caption{Thermal evolution of neo-neutron stars for selected values of the heating source $H_0=[10^{12},5\times 10^{12},10^{15}]$ erg/g/s and for an initial temperature of the atmosphere $T_0=8.7\times 10^6$ K. The observed data represents the X-ray light curve associated with the URCAs.}\label{fig:cool_crust_1} 
\end{figure}

In Fig.~\ref{fig:cool_crust_1} we show the cooling curves of neo-neutron stars resulting from the presence of the heating source given by Eq.~(\ref{eq:H}), in addition to the traditional cooling processes of neutron stars. The cooling curves are obtained self-consistently by exactly solving the full, general relativistic energy transport and balance equations as described in \cite{Schaab1996}, \cite{Page2006} and \cite{Negreiros2010}. We also show the observed data for the X-ray light curve associated with the URCAs. This allows us to identify the key factor leading to the matching of the neo-neutron star luminosity with the X-ray emission of the URCAs.

\section{Discussion and conclusions}

The major role played by the neutrino emissions from the crust of a neo-neutron star in its initial stages of the object is illustrated in Fig.~\ref{fig:cool_crust_1}. In addition, by calibrating our additional heating source at early times to $H_0 \sim 10^{12}$--$10^{15}$ erg/g/s, we find striking agreement of the luminosity obtained from the cooling of a neo-neutron stars with the prolonged ($t=10^{8}$--$10^{9}$ s) X-ray emission observed in GRB associated with an SN (see Fig.~\ref{fig:cool_crust_1} for details). This could indicate that something might be missing in our current understanding of the crust of neutron stars. It might be that, as is the case for the atmosphere, we need to develop our current models for the crust further, so as to describe properly the properties of neo-neutron stars. The traditional thermal processes taking place in the crust might be enhanced by the extremely high temperature conditions of neo-neutron star and, additional heating processes not yet studied within this context could also take place under such conditions and deserve further analysis.

Particularly interesting in this respect are the processes of $e^+e^-$ pair creation expected to occur in the interphase between the core and the crust during the neutron star formation leading to the appearance of critical fields \citep[see][for details]{ 2007IJMPD..16....1R,2008pint.conf..207R,2010JKPS...57..560R,2010AIPC.1205..143R,2010AIPC.1205..127P,2010PhR...487....1R,2011PhRvC..83d5805R,2011PhLB..701..667R,2011IJMPD..20.1789R,2011IJMPD..20.1995R,2011arXiv1107.2777R,2011arXiv1104.4062R}

It is also worth mentioning that the additional heating source needed at early times, $H_0 \sim 10^{12}$--$10^{15}$ erg/g/s (or $H_0 \sim 10^{-6}$--$10^{-3}$ MeV/Nucleon/s), is in striking agreement with the heat released from nuclear fusion reactions, radiative neutron captures, and photodisintegrations in the early stages of neutron star mergers found by \cite{2011ApJ...738L..32G,2011A&A...531A..78G}. Both fission and $\beta$-decays have also been included there: i.e neutron-induced fission, spontaneous fission, $\beta$-delayed fission, photofission, and $\beta$-delayed neutron emission.

All this suggests the exciting possibility that we are, for the first time, observing a nascent hot neutron star. This possibility alone warrants further study of this subject, so we might obtain a more concrete picture of the thermal evolution of neo-neutron stars. A proposal has been recently submitted by E.~Pian et al. to the Chandra satellite to observe whether a similar prolonged X-ray emission also exists in GRB100316D that is associated with SN2010bh \citep{pian2011}. We also encourage dedicated observations of isolated SNe in view of the similarities between URCA-1--URCA-3 and the Type Ic supernova SN 1994I \citep{2002ApJ...573L..27I} and SN 2002ap \citep{2004A&A...413..107S}.

\begin{acknowledgements}
We thank the referee for very interesting remarks.
\end{acknowledgements}



\end{document}